\documentclass[a4paper]{jpconf}
\usepackage{graphicx}
\usepackage{epsfig} 
\begin{document}
\title{ArgoNeuT, a liquid argon time projection chamber in a low energy neutrino beam}

\author{Joshua Spitz, for the ArgoNeuT Collaboration}
\address{Department of Physics, Yale University, New Haven, CT 06520, USA}
\ead{joshua.spitz@yale.edu}

\begin{abstract}
ArgoNeuT (Argon Neutrino Test), a NSF/DOE project at Fermilab, is the first LArTPC to go in a low energy neutrino beam and just the second LArTPC to go in a neutrino beam ever. ArgoNeuT sits just upstream of the on-axis MINOS near detector in the NuMI beamline, about 1~km from the target station and 100~m underground. The detector features a 47$\times$40$\times$90~cm (169~L) active volume TPC with a fully contained recirculation and purification system. Among other physics, ArgoNeuT will measure the charged-current quasi-elastic (anti-) neutrino cross section on argon at $E_{\nu}\sim3$~GeV. 
\end{abstract}

\section{Introduction}
Liquid Argon Time Projection Chambers (LArTPCs)~\cite{rubbia}
feature high-resolution three dimensional tracking and calorimetry for efficient and low background neutrino detection. An accelerator-based long baseline neutrino
oscillation experiment will require a multi-kiloton scale far detector~\cite{modular}\cite{landd}\cite{glacier}\cite{flare} for sensitivity to $\theta_{13}$ and leptonic CP-violation. LArTPCs of up to
600~tons~\cite{Icarus:2004} have been successfully built and operated but further research and design is
necessary for a multi-kiloton detector. 

In a LArTPC, the charged particles emanating from the interaction vertex ionize argon atoms as they traverse the detector. As argon is a noble element, this ionization is free to drift across the medium. Argon, with a comparatively large density, short radiation length, and high scintillation yield, is ideal for neutrino detection and event containment. Krypton and Xenon, although superficially attractive for the same reasons, are far less abundant in the atmosphere and therefore very expensive. An electric field is imposed in the active volume and the ionization tracks are drifted towards readout wire planes that are oriented at an angle with respect to one another. The ionization induces a current on the induction plane(s) and is collected by the collection plane. Combining the positional information from the wire planes and detecting the events in time allows a high-resolution three-dimensional image of the event with calorimetric information\footnote{The charge induced/collected is proportional to the amount of energy the particle deposited in the detector.} to be obtained. 
\section{ArgoNeuT}
The ArgoNeuT LArTPC is currently taking data in the on-axis Neutrinos at the Main Injector (NuMI) beamline (peaking at $E_{\nu}\sim3$~GeV) at Fermilab. ArgoNeuT instruments a total of 480 wires on an induction plane and a collection plane, oriented at an angle of $60^{\circ}$ with respect to one another. The wire spacing on both planes is 4~mm, the planes are separated by 4~mm, and the electric field is 500~V/cm in the 47$\times$40$\times$90~cm TPC. The event trigger and time of the initial neutrino interaction, $t_{0}$, are obtained from beam timing information (the NuMI beam spill window is approximately 10~$\mu s$). ArgoNeuT purifies its liquid argon through a fully contained recirculation system which includes a cryocooler and oxygen/water filters. The filter system is similar to that which is described in~\cite{filterpaper}. 
\subsection{Charged-Current Quasi-Elastic Events in ArgoNeuT}
The simple two particle event topology and dominant cross section in the relevant energy region make the Charged-Current Quasi-Elastic (CCQE\footnote{For neutrinos, $\nu_{\l}n\longrightarrow\l^{-}p$. For anti-neutrinos, $\overline{\nu}_{\l}p\longrightarrow\l^{+}n$.}) interaction the most important interaction channel for accelerator-based neutrino oscillation experiments. Current (MiniBooNE~\cite{miniboone}, MINOS~\cite{MINOS}) and future (T2K~\cite{t2k}, NOvA~\cite{nova}, wide-band-beam to DUSEL~\cite{dusel}, ...) accelerator-based neutrino oscillation experiments search for the appearance of electron neutrinos largely via the $\nu_{e}$ CCQE interaction $\nu_{e}n\longrightarrow e^{-}p$. The $\nu_{\mu}$ CCQE interaction is used to measure the muon-neutrino flux, constraining the expected $\nu_{e}$ flux at the near and far detectors. The anti-neutrino analog interactions are important in the search for CP violation in the lepton sector (e.g. $P[\nu_{\mu}\rightarrow \nu_{e}]\neq P[\overline{\nu}_{\mu}\rightarrow\overline{\nu}_{e}]$). The CCQE cross section is known with 20-30\% uncertainty over most relevant energies and only one (preliminary) measurement of the $\nu_{\mu}$ CCQE cross section on argon has ever been taken (at $<E_{\nu}>\sim28$~GeV)~\cite{icaruswanf}.  
\begin{figure}[h!]
\centering
\begin{tabular}{c c}
\hspace{-.3cm}
\epsfig{file=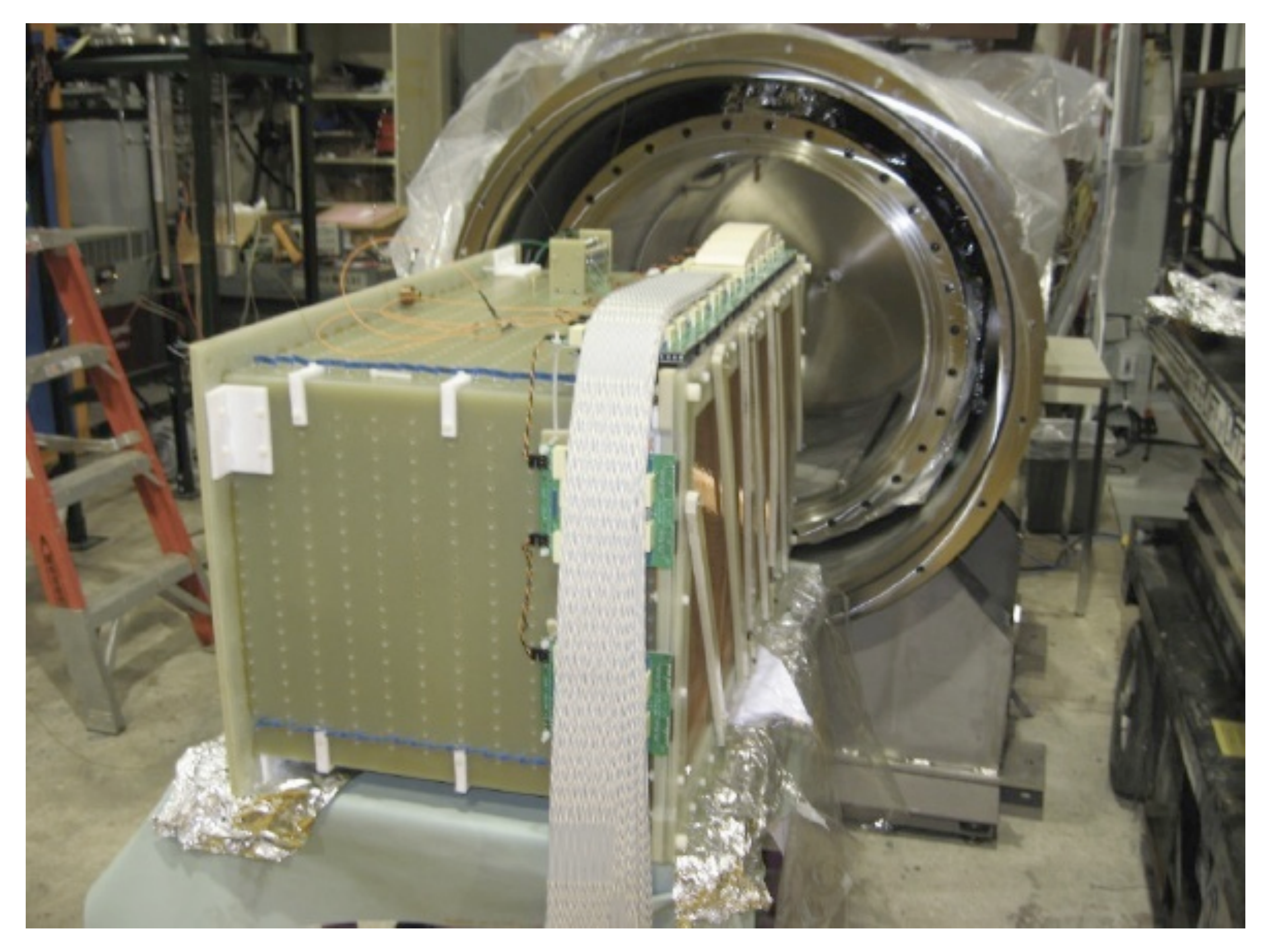,width=3.1in,angle=0}
&
\hspace{-0.1in}
\epsfig{file=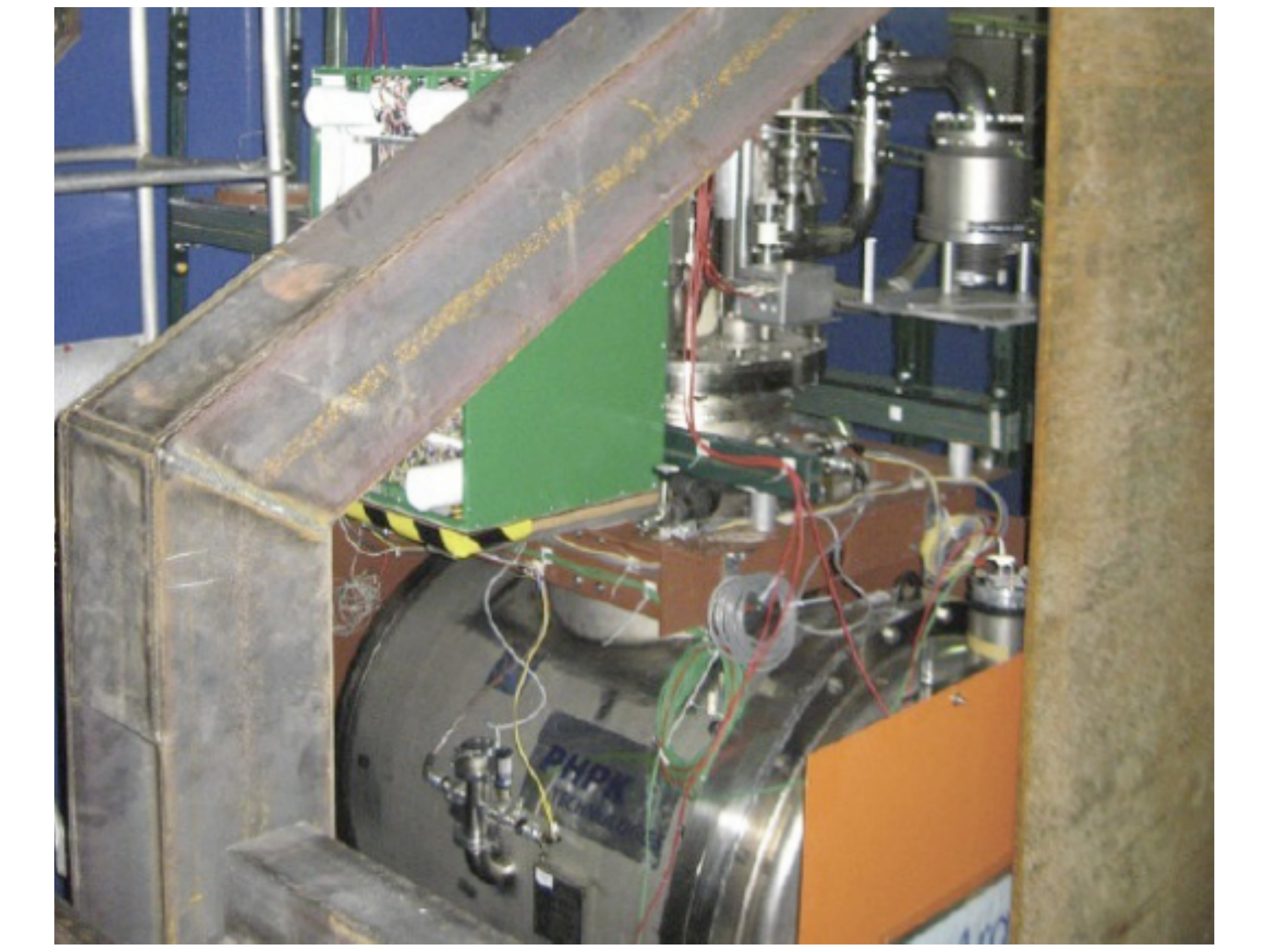,width=3.1in,angle=0}
\end{tabular}
\vspace{0.0in}
\caption{Left: The fully instrumented TPC being inserted into the ArgoNeuT inner cryostat. The cryostat is composed of an inner, liquid argon layer and an outer, vacuum-jacket layer. Right: The ArgoNeuT cryostat in the NuMI beamline.}
\label{argoneutpic}
\end{figure}

ArgoNeuT will detect $\sim$1230 anti-neutrino CCQE events/180 days with 54\% proton containment inside the active volume. ArgoNeuT will employ the downstream MINOS near detector for sign determination and muon energy reconstruction as it is usually necessary to range out the muon in order to know its energy precisely. Larger LArTPCs can reconstruct the muon's energy via multiple scattering if the muon travels longer than about 1~m~\cite{multiplescattering}. However, a muon will rarely travel this far in ArgoNeuT. Note that in 47\% of
CCQE events, the proton is contained inside ArgoNeuT and the muon enters the MINOS near detector. ArgoNeuT will measure the CCQE interaction rate as a function of neutrino energy and determine cross section using the well known NuMI neutrino flux\footnote{The NuMI neutrino flux at ArgoNeuT is known with 5-10\% uncertainty.}.
\begin{figure}
\begin{center}
\includegraphics[width=21pc]{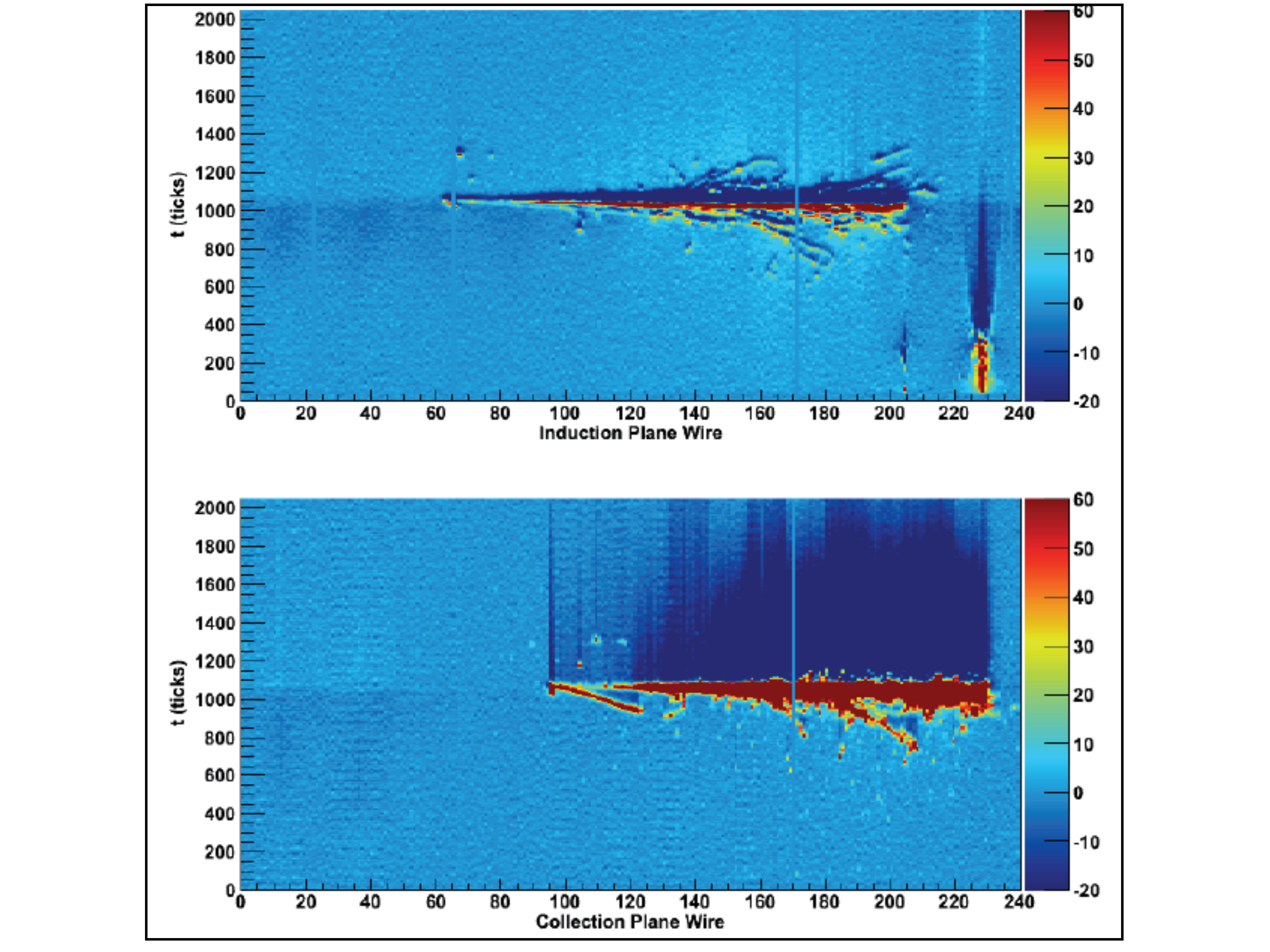}
\end{center}
\vspace{-.2in}
\caption{\label{nue}A $\nu_{e}$ charged-current candidate event an seen in the induction and collection views. Please note that this event image is raw data and that the ``$\nu_{e}$ charged-current" interpretation is preliminary.}
\end{figure}
\section{Outlook}
ArgoNeuT was installed in the NuMI beamline in early 2009 and filled with liquid argon in May 2009. The first neutrino events were observed shortly thereafter. ArgoNeuT ran in neutrino-commissioning-mode for $\sim$1 month up until the planned 3 month beam shutdown on June 14, 2009 and the physics run began on September 14, 2009 (ending on March 1, 2010). A recent neutrino event in ArgoNeuT can be seen in Figure~\ref{nue}. 

\ack
The author would like to acknowledge the support staff at Fermilab for their invaluable contributions to the planning and construction of ArgoNeuT. This work is supported by the Department of Energy and the National Science Foundation.

\section*{References}
\bibliographystyle{iopart-num}
\bibliography{./bibfile}
\end{document}